\documentstyle[12pt]{article}
\catcode`\@=11

\@addtoreset{equation}{section}
\def\theequation{\arabic{equation}}

\catcode`\@=11

\def\section{\@startsection{section}{1}{\z@}{3.5ex plus 1ex minus
   .2ex}{2.3ex plus .2ex}{\large\bf}}

\newskip\humongous \humongous=0pt plus 1000pt minus 1000pt

\newif\ifdtup

\def\eqnarray{\let\@currentlabel=\theequation\refstepcounter{equation}
    \global\@eqnswtrue
    \global\@eqcnt\z@\tabskip\@centering\let\\=\@eqncr
    $$\halign to \displaywidth\bgroup\@eqnsel\hskip\@centering
      $\displaystyle\tabskip\z@{##}$&\global\@eqcnt\@ne
       \hfil${{}##{}}$\hfil
      &\global\@eqcnt\tw@ $\displaystyle\tabskip\z@{##}$\hfil
       \tabskip\@centering&\llap{##}\tabskip\z@\cr}
\def\lefteqn#1{\hbox to 4\arraycolsep{$\displaystyle #1$\hss}}
\def\thesection{\arabic{section}.}

\def\appendix{\setcounter{section}{0}
        \def\thesection{Appendix.}
        \def\theequation{\Alph{section}.\arabic{equation}}}

\long\def\@makefntext#1{\parindent 0cm\noindent
\hbox to 1em{\hss$^{\@thefnmark}$}#1}
\def\IR{{\hbox{{\rm I}\kern-.2em\hbox{\rm R}}}}
\def\IH{{\hbox{{\rm I}\kern-.2em\hbox{\rm H}}}}
\def\IC{{\ \hbox{{\rm I}\kern-.6em\hbox{\bf C}}}}
\def\IZ{{\hbox{{\rm Z}\kern-.4em\hbox{\rm Z}}}}
\def\rref#1{(\ref{#1})}
\newcommand{\beq}{\begin{equation}}
\newcommand{\eeq}{\end{equation}}
\newcommand{\Ann}[1]{{\sl Ann.~Phys.}~{\bf #1}}

\newcommand{\PRD}[1]{{\sl Phys.~Rev.}~{\bf D#1}}

\begin{document}
%
%
%
%
\def\citen#1{%
\edef\@tempa{\@ignspaftercomma,#1, \@end, }
\edef\@tempa{\expandafter\@ignendcommas\@tempa\@end}%
\if@filesw \immediate \write \@auxout {\string \citation {\@tempa}}\fi
\@tempcntb\m@ne \let\@h@ld\relax \let\@citea\@empty
\@for \@citeb:=\@tempa\do {\@cmpresscites}%
\@h@ld}
%
\def\@ignspaftercomma#1, {\ifx\@end#1\@empty\else
   #1,\expandafter\@ignspaftercomma\fi}
\def\@ignendcommas,#1,\@end{#1}
%
%
\def\@cmpresscites{%
 \expandafter\let \expandafter\@B@citeB \csname b@\@citeb \endcsname
 \ifx\@B@citeB\relax 
    \@h@ld\@citea\@tempcntb\m@ne{\bf ?}%
    \@warning {Citation `\@citeb ' on page \thepage \space undefined}%
 \else
    \@tempcnta\@tempcntb \advance\@tempcnta\@ne
    \setbox\z@\hbox\bgroup 
    \ifnum\z@<0\@B@citeB \relax
       \egroup \@tempcntb\@B@citeB \relax
       \else \egroup \@tempcntb\m@ne \fi
    \ifnum\@tempcnta=\@tempcntb 
       \ifx\@h@ld\relax 
          \edef \@h@ld{\@citea\@B@citeB}%
       \else 
          \edef\@h@ld{\hbox{--}\penalty\@highpenalty \@B@citeB}%
       \fi
    \else   
       \@h@ld \@citea \@B@citeB \let\@h@ld\relax
 \fi\fi%
 \let\@citea\@citepunct
}
%
\def\@citepunct{,\penalty\@highpenalty\hskip.13em plus.1em minus.1em}%
%
%
\def\@citex[#1]#2{\@cite{\citen{#2}}{#1}}%
%
%
\def\@cite#1#2{\leavevmode\unskip
  \ifnum\lastpenalty=\z@ \penalty\@highpenalty \fi 
  \ [{\multiply\@highpenalty 3 #1
      \if@tempswa,\penalty\@highpenalty\ #2\fi 
    }]\spacefactor\@m}
\let\nocitecount\relax  
%
\begin{titlepage}
\vspace{.5in}
\begin{flushright}
IASSNS-HEP-93/84\\
UCD-93-34\\
gr-qc/9312002\\
November 1993\\
\end{flushright}
\vspace{.5in}
\begin{center}
{\Large\bf
The Off-Shell Black Hole}\\
\vspace{.4in}
S{\sc teven}~C{\sc arlip}\footnote{\it email: carlip@dirac.ucdavis.edu}\\
       {\small\it Department of Physics}\\
       {\small\it University of California}\\
       {\small\it Davis, CA 95616}\\{\small\it USA}\\
\vspace{1ex}
{\small and}\\
\vspace{1ex}
C{\sc laudio}~T{\sc eitelboim}\footnote{\it email: cecsphy@lascar.puc.cl}\\
       {\small\it Centro de Estudios Cientificos de Santiago}\\
       {\small\it Casilla 16443, Santiago 9}\\{\small\it Chile}\\
       {\small\it and}\\
       {\small\it Institute for Advanced Study}\\
       {\small\it Olden Lane, Princeton, NJ 08540}\\{\small\it USA}\\
\end{center}

\vspace{.5in}
\begin{center}
\begin{minipage}{5in}
\begin{center}
{\large\bf Abstract}
\end{center}
{\small
The standard (Euclidean) action principle for the gravitational field
implies that for spacetimes with black hole topology, the opening angle
at the horizon and the horizon area are canonical conjugates.  It is
shown that the opening angle bears the same relation to the horizon
area that the time separation bears to the mass at infinity.  The
dependence of the wave function on this new degree of freedom is
governed by an extended Wheeler-DeWitt equation.  Summing over all
horizon areas yields the black hole entropy.
}
\end{minipage}
\end{center}
\end{titlepage}
\addtocounter{footnote}{-2}

It has long been known that when
space is not closed, the wave functional of the gravitational field
possesses an extra argument in addition to the intrinsic geometry of
a spatial section.  This additional argument is the time separation
at spatial infinity, which is conjugate to the total mass, given by
a surface integral.  Thus the wave functional $\psi = \psi[T,{}^{(3)}g]$
obeys a Schr\"odinger equation
\beq
{\hbar\over i}{\partial\psi\over\partial T} + M\psi = 0
\label{a1}
\eeq
in addition to the constraints
\beq
{\cal H}_\mu \psi = 0 .
\label{a2}
\eeq
Here $M$ is the ADM mass, a surface integral over the boundary of space
at infinity (a two-dimensional surface in four spacetime dimensions).
Equations \rref{a1} and \rref{a2} are not independent, and are more
appropriately written in the form \cite{CT}
\beq
{\hbar\over i}\delta\psi + \left[ \int\delta t N^\mu{\cal H}_\mu\, d^3x
  + \delta T\cdot M \right] \psi = 0
\label{a3}
\eeq
for any $N^\mu = (N, N^i)$ obeying the asymptotic conditions
\beq
N(t_2-t_1)\rightarrow T , \qquad N^i\rightarrow 0
\label{a4}
\eeq
at $r\rightarrow\infty$.  (See \cite{RT} for a precise statement of these
conditions.)  If spacetime has the topology $\IR^4$, equation \rref{a3}
extends the Wheeler-DeWitt equation.

The purpose of this note is to point out that for black hole topologies,
the wave function $\psi$ acquires an extra argument $\Theta$, which is a
sort of ``dimensionless internal time'' associated with the horizon.
The dependence of $\psi$ on $\Theta$ is governed by an equation
analogous to \rref{a1},
\beq
{\hbar\over i}{\partial\psi\over\partial \Theta} - A\psi = 0 ,
\label{a5}
\eeq
where $A$ is a surface integral at the horizon, the horizon area.
Equation \rref{a3} is replaced by
\beq
{\hbar\over i}\delta\psi + \left[ -\delta\Theta\cdot A
  + \int\delta\xi^\mu{\cal H}_\mu\, d^3x
  + \delta T\cdot M \right] \psi = 0 .
\label{a6}
\eeq
The ``off-shell'' meaning of the term horizon and the analog of \rref{a4}
for $\Theta$ will be given below.

The transition amplitude now depends on an additional argument,
\beq
K = K[{}^{(3)}{\cal G}_2,{}^{(3)}{\cal G}_1; A_2,A_1; M_2,M_1]
  = \delta(A_2-A_1)\delta(M_2-M_1)
  K[{}^{(3)}{\cal G}_2,{}^{(3)}{\cal G}_1;A_2;M_2] .
\label{a7a}
\eeq
If one decrees that ``the horizon area $A$ is not observable,'' one may
introduce a reduced amplitude
\beq
K[{}^{(3)}{\cal G}_2,{}^{(3)}{\cal G}_1;M_2] =
  \int dA\,K[{}^{(3)}{\cal G}_2,{}^{(3)}{\cal G}_1;A;M] ,
\label{a7b}
\eeq
which may be rewritten as
\beq
K[{}^{(3)}{\cal G}_2,{}^{(3)}{\cal G}_1;M]
  = \tilde K[{}^{(3)}{\cal G}_2,{}^{(3)}{\cal G}_1;\Theta_E;M]
  \Bigl|_{\Theta_E=2\pi} \ ,
\label{a8}
\eeq
where $\tilde K[{}^{(3)}{\cal G}_2,{}^{(3)}{\cal G}_1;\Theta_E;M]$ is
a Laplace transform with respect to $A$ of the amplitude appearing
on the right-hand side of \rref{a7a}.  (The shift by $2\pi$ in
the argument of the Laplace transform will be explained below.)
The trace of \rref{a8} in ${}^{(3)}{\cal G}_2$ and ${}^{(3)}{\cal G}_1$
yields the exponential of the black hole entropy.

The analysis leading to the preceding statements goes as
follows.\footnote{For definiteness, we will work in four spacetime
dimensions.  The reasoning remains valid in any dimension, however,
and in particular applies to the (2+1)-dimensional black hole, where
many of these features were first discovered.  A detailed treatment
of the (2+1)-dimensional case will appear elsewhere \cite{SC+CT}.
Variations of the conical singularity at the horizon have also been
considered by Susskind \cite{LS}.}

We start from the Euclidean point of view.  The spacetimes admitted
in the action principle will have the topology $\IR^2\!\times\!S^{d-2}$
(``one black hole sector'').  We introduce a system of polar coordinates
$r$, $\tau$ in $\IR^2$, with an arbitrary origin $r=r_+$ for the radial
coordinate.  In the semiclassical approximation, it is useful to take
the origin to be the horizon of the black hole and the angle $\tau$ to
be the Killing time.  By abuse of language, we will call $r_+$ the
horizon even away from the extremum (``off shell'').

Although it is unnecessarily complicated, we will conform to standard
practice and use ``Schwarzschild coordinates'' near $r_+$.  That is,
we write the generic Euclidean metric as
\beq
ds^2_E = N^2(r)d\tau^2 + N^{-2}(r)dr^2 + \gamma_{mn}(r,x^p)dx^mdx^n
\label{a9}
\eeq
up to terms of order $O(r-r_+)$, with
\beq
(\tau_2-\tau_1)N^2 = 2\Theta_E(r-r_+) + O(r-r_+)^2 .
\label{a10}
\eeq
Here the $x^m$ are coordinates on the two-sphere $S^2$.  The parameter
$\Theta_E$ is the total proper angle (proper length divided by proper
radius) of an arc of very small radius and coordinate angular opening
$\tau_2-\tau_1$.  For this reason it will be called the ``opening
angle.''  If one identifies the surfaces $\tau=\tau_1$ and $\tau=\tau_2$,
thus considering a disk in $\IR^2$, then the deficit angle $2\pi-\Theta_E$
is the strength of a conical singularity in $\IR^2$ at $r_+$.  For the
moment, we assume for simplicity that $\Theta_E$ is independent of $x^m$;
we shall see below that this restriction may be lifted without changing
the conclusions.  It is important to emphasize that no a priori relation
between $\Theta_E$ and $N(\infty)$ is assumed.

Equation \rref{a10} is the Euclidean analog of \rref{a4} for $\Theta$.
Note that if we continue the metric \rref{a9} to Minkowskian signature
by setting $\tau = it$, we find
\beq
ds^2 = -N^2dt^2 + N^{-2}dr^2 + \gamma_{mn}(r,x^p)dx^mdx^n
\label{a11}
\eeq
with
\beq
i(t_2-t_1)N^2 = 2\Theta_E(r-r_+) + O(r-r_+)^2 .
\label{a12}
\eeq
We therefore learn that the opening angle must also be rotated,
\beq
\Theta_E = i\Theta .
\label{a13}
\eeq

Now, it was shown in reference \cite{BTZ} by a geometrical argument
based on the Gauss-Bonnet theorem that the covariant Hilbert action
\beq
I_H = {1\over2}\int_M\sqrt{g}R\,d^4x
    - \int_{\partial M} \sqrt{h}K\, d^3x
\label{a14}
\eeq
for $\IR^2\!\times\!S^{d-2}$ and the canonical action
\beq
I_{\hbox{\scriptsize\it can}} =
    \int_M\left( \pi^{ij}{\partial g_{ij}\over\partial \tau}
    - N^\mu{\cal H}_\mu \right)d^4x
\label{a15}
\eeq
that uses the polar angle in $\IR^2$ as time are related by
\beq
I_H = 2\pi A + I_{\hbox{\scriptsize\it can}} + B_\infty ,
\label{a16}
\eeq
where $B_\infty$ is a local boundary term at spatial infinity.
The action \rref{a16} is suitable for fixing the intrinsic geometry
of the boundary at infinity, for either the full manifold
$\IR^2\!\times\!S^{d-2}$ or the wedge $\tau_1\le\tau\le\tau_2$.
If we instead wish to implement black hole
boundary conditions, fixing $N(\infty)$ with a prescribed rate of
fall-off for $N-N(\infty)$ and holding the angular momentum and gauge
charges fixed \cite{RT}, then the term $B_\infty$ (which diverges)
must be supplemented by a further surface term, leading to an action
\beq
I = 2\pi A + I_{\hbox{\scriptsize\it can}} - T_EM ,
\label{a17}
\eeq
where $T_E$ is the Euclidean time separation at infinity.  The fields
held fixed in the action \rref{a17} are the following: (i)~the
three-geometries of the hypersurfaces at $\tau=\tau_1$ and $\tau=\tau_2$
for $r>r_+$; (ii)~the two-geometry at the horizon; and (iii)~the
asymptotic time separation $T_E$.

The variation of \rref{a17} is given by
\beq
\delta I = (2\pi-\Theta_E)\delta A - \delta T_E M
  + \int \pi^{ij}\delta g_{ij}\,d^3x
  \left|^{\raise5pt\hbox{$\scriptstyle2$}}_{\lower5pt\hbox{$\scriptstyle1$}}
  \right. + \hbox{\it terms vanishing on shell} .
\label{a18}
\eeq
The contribution $-\Theta_E\delta A$ to the variation comes technically
from an integration by parts in the variation of ${\cal H}_\perp$ in
$I_{\hbox{\scriptsize\it can}}$.  A more geometric way to understand
this term is to recall that the action \rref{a17} differs from the scalar
curvature by a boundary term at infinity, so the contribution to the
action of a neighborhood of the horizon is just the integrated scalar
curvature.  This immediately brings in the deficit angle $2\pi-\Theta_E$,
a measure of the curvature in $\IR^2$ per unit area in $S^2$.

It follows from \rref{a18} that in addition to the ordinary degrees of
freedom (i.e., those present when the topology is $\IR^4$) there is
one additional dynamical variable, $\Theta_E$, conjugate to the area $A$.
The action takes a more symmetric form if one passes to a representation
suitable for fixing $\Theta_E$ instead of $A$.  This can be achieved by
subtracting $(2\pi-\Theta_E)A$ from \rref{a17}, yielding
\beq
I^\prime = \Theta_E A + I_{\hbox{\scriptsize\it can}} - T_EM .
\label{a19}
\eeq
In this form the symmetry between $\Theta$ and $T$ is manifest, and
equations \rref{a5} and \rref{a6} are established.  (Recall that
$\Theta_E = i\Theta$ and $T_E = iT$.)

One may further allow for a dependence of $\Theta_E$ on the coordinates
$x^m$ of the two-sphere at $r_+$.  The first term on the right-hand side
of \rref{a19} must then be replaced by
$$\int \Theta_E(x) \gamma^{1/2}(x) d^2x .$$
Thus $\Theta_E(x)$ is canonically conjugate to the local measure of area
on the horizon, $\gamma^{1/2}(x)$, and equation \rref{a5} may be
obtained as the integral over the horizon of the ``many-time equation''
\beq
{\hbar\over i}{\delta\psi\over\delta\Theta(x)} - \gamma^{1/2}(x)\psi = 0 .
\label{ax}
\eeq
The reduced amplitude is obtained by summing over all $\gamma^{1/2}(x)$.
This amounts to setting $\Theta_E(x) =2\pi$ for all $x$, so \rref{a8}
remains valid.

In the semiclassical approximation, the trace of \rref{a8} is given by the
exponential of the action \rref{a19} evaluated on the black hole solution. If
one takes $\tau$ to be the Killing time, then $I_{\hbox{\scriptsize\it can}}$
has the value zero, and one finds the standard result
\beq
S = (8\pi G\hbar)^{-1} 2\pi A_{\hbox{\scriptsize\it horizon}}
\label{a20}
\eeq
for the entropy.  We have restored the universal constants, in order to
make explicit the geometrical nature of the factor multiplying the area in
\rref{a20} as the opening angle.  In this sense, the most natural choice
of units is $G=(8\pi)^{-1}$ rather than $G=1$.

The above analysis shows that one may regard the black hole entropy as
arising from summing over all horizon geometries.  We still lack
a ``microscopic'' explanation for the exponential weight in the integration
measure for the surface degrees of freedom, or equivalently for the
$\hbar^{-1}$ dependence in \rref{a20}.  Note, however, that the factor
multiplying the area in (\ref{a20}) comes from the action of a small disk
in $\IR^2$.  This would suggest that the entropy per unit area may arise
from counting the geometries in such a ``thickened horizon.''

\vspace{1.5ex}
\begin{flushleft}
\large\bf Acknowledgements
\end{flushleft}

We would like to thank M\'aximo Ba\~nados and Jorge Zanelli for enlightening
discussions.  This work was partially supported by grants 0862/91 and
193.1910/93 from FONDECYT (Chile), by institutional support to the Centro
de Estudios Cientificos de Santiago provided by SAREC (Sweden) and a group
of Chilean private companies (COPEC, CMPC, ENERSIS, CGEI).  This research
was also sponsored by CAP, IBM and Xerox de Chile.  S.C.\ was supported in
part by U.S.\ Department of Energy grant DE-FG03-91ER40674 and National
Science Foundation National Young Investigator award PHY-93-57203.

\end{document}